\documentclass[12pt]{article}
\usepackage[dvips]{graphicx}
\usepackage{feynmf}
\usepackage{amsmath,amssymb}
\usepackage{epsfig}
\usepackage{subfigure}
\usepackage{amsmath}
\usepackage{amsfonts}
\usepackage{amssymb}
\usepackage{url}
\usepackage{hyperref}
\newcommand{\be}{\begin{equation}}
\newcommand{\ee}{\end{equation}}
\newcommand{\bea}{\begin{eqnarray}}
\newcommand{\eea}{\end{eqnarray}}

\begin{document}
\begin{titlepage}
\begin{flushright}
\end{flushright}
\vspace{4\baselineskip}
\begin{center}
{\Large\bf 
Non-thermal Leptogenesis in a simple 5D SO(10) GUT          
}
\end{center}
\vspace{1cm}
\begin{center}
{\large Takeshi Fukuyama $^{a,}$
\footnote{\tt E-mail:fukuyama@se.ritsumei.ac.jp}
and Nobuchika Okada $^{b,}$
\footnote{\tt E-mail:okadan@ua.edu}} 
\end{center}
\vspace{0.2cm}
\begin{center}
${}^{a}$ {\small \it Department of Physics and R-GIRO, 
 Ritsumeikan University, Kusatsu, Shiga, 525-8577, Japan}\\
${}^{b}$ {\small \it  
Department of Physics and Astronomy, 
University of Alabama, Tuscaloosa, AL 35487, USA}\\
\medskip
\vskip 10mm
\end{center}
\vskip 10mm

\begin{abstract} 
We discuss the non-thermal leptogenesis 
 in the scheme of 5D orbifold SO(10) GUT 
 with the smooth hybrid inflation. 
With an unambiguously determined Dirac Yukawa couplings 
 and an assumption for the neutrino mixing matrix 
 of the tri-bimaximal from, we analyze baryon asymmetry 
 of the universe via non-thermal leptogenesis 
 in two typical cases for the light neutrino mass spectrum, 
 the normal and inverted hierarchical cases. 
The resultant baryon asymmetry  is obtained as a function 
 of the lightest mass eigenvalue of the light neutrinos, 
 and we find that a suitable amount of baryon asymmetry 
 of the universe can be produced in the normal hierarchical case,
 while in the inverted hierarchical case 
 the baryon asymmetry is too small to be consistent  
 with the observation. 

\end{abstract}
\end{titlepage}


%
The so-called renormalizable minimal SO(10) GUT model 
 has been paid a particular attention, 
 where two Higgs multiplets $\{{\bf 10} \oplus {\bf \overline{126}}\}$ 
 are utilized for the Yukawa couplings with matters 
 ${\bf 16}_i$ ($i=1,2,3$ is the generation index) 
 \cite{B-M, Fuku1, Fuku2}. 
A remarkable feature of the model is its high predictive power 
 for the neutrino oscillation parameters, in reproducing 
 charged fermion masses and mixing angles. 
The unambiguously determined Yukawa couplings play a crucial role 
 for predictions of the model in other phenomena, 
 such as the lepton flavor violation \cite{Fuku3} 
 and proton decay \cite{FKO-proton}. 
The Higgs superpotential of the model has been investigated 
 and the explicit pattern of the SO(10) gauge symmetry to 
 the standard model one has been shown \cite{Fuku4, Goran}. 
On the other hand, the explicit representation of intermediate 
 energy scales revealed in these papers gives rise to 
 the deviation of gauge coupling unification \cite{Bertolini}. 
Also the minimal SO(10) model likely predict too short proton
 lifetime and has some deviation from the precise measurements 
 of the neutrino oscillation data \cite{Kamland} 
 (see however \cite{bajc}).

In order to remedy these problems, we have argued SO(10) GUT 
 in the context of the orbifold GUT 
 \cite{Kawamura, Altarelli, Koba, Hall, Hebecker} 
 and proposed a simple supersymmetric (SUSY) SO(10) model 
 in 5D \cite{F-O1}. 
In this model, the SO(10) symmetry in 5D is broken by 
 orbifold boundary conditions to the Pati-Salam (PS) symmetry 
 SU(4)$_C\times$ SU(2)$_L\times$SU(2)$_R$. 
All matter and Higgs multiplets reside only 
 on a brane (PS brane) where the PS gauge symmetry is manifest, 
 so that low energy effective description of this model 
 is nothing but the PS model in 4D with a special set of 
 matter and Higgs multiplets. 
At energies higher than the compactification scale, 
 the Kaluza-Klein (KK) modes of the bulk SO(10) gauge multiplet 
 are involved in the particle contents and in fact, the gauge coupling 
 unification was shown to be successfully realized 
 by incorporating the KK mode threshold corrections 
 into the gauge coupling running \cite{F-O1}. 
The unification scale ($M_{\rm GUT}$) and 
 the compactification scale ($M_c$)
 which was set to be the same as the PS symmetry breaking 
 scale ($v_{\rm PS}$) for simplicity were found to be 
 $M_{\rm GUT} = 4.6 \times 10^{17}$ GeV and 
 $M_c=v_{\rm PS} = 1.6 \times 10^{16}$ GeV, respectively.

More recently, it has been shown \cite{F-O2} that 
 this orbifold GUT model is applicable to the smooth 
 hybrid inflation \cite{SmoothHI}. 
Interestingly, this inflation model can fit the WMAP data \cite{WMAP} 
 very well by utilizing the PS breaking scale ($v_{\rm PS}$) and 
 the gauge coupling unification scale predicted independently 
 of cosmological considerations. 
Another cosmological issue, the dark matter candidate of the model 
 has been investigated in \cite{F-O3}. 
In the paper, the sparticle mass spectrum is calculated 
 in the context of the gaugino mediated supersymmetry breaking 
 \cite{gMSB} which can be naturally incorporated in the model 
 and it has shown that the neutralino LSP as 
 the dark matter candidate can be realized 
 when the compactification scale is taken to be  
 slightly bigger than the PS symmetry breaking scale, 
 while keeping the successful gauge coupling unification.

In the present paper, we apply our model to the leptogenesis 
 scenario for creating the baryon asymmetry of the universe. 
In order to produce a suitable amount of the baryon asymmetry 
 of the universe in the thermal leptogenesis scenario \cite{F-Y}, 
 the scale of right-handed (scalar) neutrino masses should 
 be grater than $10^9$ GeV \cite{LowerBound} and hence the reheating 
 temperature after inflation should also be beyond this scale. 
However, in supersymmetric models, 
 the reheating temperature is severely constrained  
 by Big Bang Nucleosynthesis (BBN) 
 to be $T_R \lesssim 10^6$ GeV \cite{BBN} 
 (gravitino problem \cite{gravitinoP}), 
 and the conventional thermal leptogenesis scenario cannot work.

In this case, we consider the so-called non-thermal leptogenesis 
 \cite{NT-LG} in which the right-handed (scalar) neutrinos 
 are non-thermally produced by the decay of inflaton and 
 their decays can produce a suitable amount of baryon asymmetry 
 of the universe even if the reheating temperature is low. 
We adopt the non-thermal leptogenesis to our hybrid inflation 
 scenario \cite{F-O2} and show that the non-thermal leptogenesis 
 is successful with a suitable choice of the model parameters 
 which are consistent with the results in the previous works 
 \cite{F-O1, F-O2, F-O3}.

Let us begin with a brief review of the orbifold SO(10) GUT 
 proposed in Ref.~\cite{F-O1}. 
The model is described in 5D and the 5th dimension is compactified 
 on the orbifold $S^1/{Z_2 \times Z_2^\prime}$ 
 \cite{Kawamura, Altarelli, Hall}. 
A circle $S^1$ with radius $R$ is divided by 
 a $Z_2$ orbifold transformation $y \to -y$ 
 ($y$ is the fifth dimensional coordinate $ 0 \leq y < 2 \pi R$)
 and this segment is further divided by a $Z_2^\prime$ transformation 
 $y^\prime \to -y^\prime $ with $y^\prime = y + \pi R/2$. 
There are two inequivalent orbifold fixed points at $y=0$ and $y=\pi R/2$. 
Under this orbifold compactification, a general bulk wave function 
 is classified with respect to its parities,  
 $P=\pm$ and $P^\prime=\pm$, under $Z_2$ and $Z_2^\prime$, respectively.

Assigning suitable parities ($P,P^\prime $) to 
 the bulk SO(10) gauge multiplet \cite{F-O1}, 
 only the PS gauge multiplet has zero-mode 
 and the bulk 5D N=1 SUSY SO(10) gauge symmetry is broken 
 to 4D N=1 supersymmetric PS gauge symmetry. 
All vector multiplets has wave functions on the brane 
 at $y=0$, SO(10) gauge symmetry is respected there, 
 while only the PS symmetry is on the brane at $y=\pi R/2$ (PS brane). 

\begin{table}[h]
{\begin{center}
\begin{tabular}{|c|c|}
\hline
& brane at $y=\pi R/2$ \\ 
\hline
& \\
Matter Multiplets & $\psi_i=F_{Li} \oplus F_{Ri}^c \quad (i=1,2,3)$ \\
 & \\
\hline
 & \\
Higgs Multiplets & 
$({\bf 1},{\bf 2},{\bf 2})_H$,  
$({\bf 1},{\bf 2},{\bf 2})'_H$,
$({\bf 15},{\bf 1},{\bf 1})_H$,
$({\bf 6},{\bf 1},{\bf 1})_H$ \\  & 
$({\bf 4},{\bf 1},{\bf 2})_H$, 
$(\overline{{\bf 4}},{\bf 1},{\bf 2})_H$, 
$({\bf 4},{\bf 2},{\bf 1})_H$, 
$(\overline{{\bf 4}},{\bf 2},{\bf 1})_H$  \\
& \\
\hline
\end{tabular}
\end{center}}
\caption{
Particle contents on the PS brane. 
$F_{Li}$ and $F_{Ri}^c$ are matter multiplets 
 of $i$-th generation in $({\bf 4, 2, 1})$ and $({\bf \bar{4}, 1, 2})$ 
 representations, respectively. 
}
\end{table}

We place all the matter and Higgs multiplets on the PS brane, 
 where only the PS symmetry is manifest, 
 so that the particle contents are in the representation 
 under the PS gauge symmetry, not necessary to be 
 in SO(10) representation. 
For a different setup, see \cite{Raby}. 
The matter and Higgs in our model is listed in Table~1. 
For later conveniences, let us introduce the following notations: 
\bea
H_1&=&({\bf 1},{\bf 2},{\bf 2})_H, ~H_1^{\prime}=({\bf 1},{\bf 2},{\bf 2})'_H,
\nonumber \\
H_6&=&({\bf 6},{\bf 1},{\bf 1})_H, ~H_{15}=({\bf 15},{\bf 1},{\bf 1})_H,
\nonumber \\ 
H_L&=&({\bf 4},{\bf 2},{\bf 1})_H,
~\overline{H_L} =(\overline{{\bf 4}},{\bf 2},{\bf 1})_H,  
\nonumber \\
\phi&=&({\bf 4},{\bf 1},{\bf 2})_H,
~\bar{\phi}=(\overline{{\bf 4}},{\bf 1},{\bf 2})_H.
\eea

Superpotential relevant for fermion masses is given by%
\footnote{
For simplicity, we have introduced only minimal terms 
 necessary for reproducing observed fermion mass matrices. 
}
\bea
W_Y&=& Y_{1}^{ij} F_{Li} F_{Rj}^c H_1 
+\frac{Y_{15}^{ij}}{M_5} F_{Li} F_{Rj}^c 
 \left(H_1^{\prime} H_{15} \right) \nonumber\\ 
&+&\frac{Y_R^{ij}}{M_5} F_{Ri}^c F_{Rj}^c 
 \left(\phi \phi \right),   
\label{Yukawa}
\eea 
where $M_5$ is the 5D Planck mass. 
The product, $H_1^{\prime} H_{15}$, effectively works 
 as $({\bf 15},{\bf 2},{\bf 2})_H$, 
 while $\phi \phi$ effectively works as $({\bf 10},{\bf 1},{\bf 3})$, 
 and is responsible for the right-handed Majorana neutrino masses. 
Assuming appropriate VEVs for Higgs multiplets, 
 fermion mass matrices are obtained, 
 which we parameterize as the following form \cite{F-O1}: 
\begin{eqnarray}
 M_u &=& c_{10} M_{1,2,2}+ c_{15} M_{15,2,2} \; , 
 \nonumber \\
 M_d &=& M_{1,2,2} + M_{15,2,2} \; ,   
 \nonumber \\
 M_D &=& c_{10} M_{1,2,2} - 3 c_{15} M_{15,2,2} \; , 
 \nonumber \\
 M_e &=& M_{1,2,2} - 3 M_{15.2,2} \; , 
 \nonumber \\
 M_R &=& c_R M_{10,1,3} \; . 
\label{massmatrix}
\end{eqnarray}  
Here, $M_u,M_d,M_D$ and $M_e$ are the mass matrices of 
 up and down type quarks, Dirac neutrino 
 and charged lepton, respectively, 
 while $M_R$ is right-handed Majorana neutrino mass matrix.

The following two points should be remarked: \\
\noindent 
1. The combination of two mass matrices of $M_{1,2,2}$ and 
 $M_{15,2,2}$ among $M_u,M_d,M_D$, and $M_e$ in the PS symmetry 
 is the same as that of $M_{10}$ and $M_{126}$ 
 in the minimal SO(10) model (see \cite{Fuku2} for notation) 
 and, therefore, the procedure for fitting the realistic 
 Dirac fermion mass matrices is the same as 
 in the minimal SO(10) model.\\ 
\noindent 
On the other hand, \\
\noindent 
2. $M_R$ is fully independent on the above 
 four Dirac Fermion mass matrices in the PS group, 
 whereas  in the minimal SO(10) model 
 it is described by $M_{126}$ and not independent. 
This fact enables us to improve the precise data fitting 
 on the neutrino oscillation parameters.

Now we discuss the smooth hybrid inflation model \cite{SmoothHI} 
 in the context of the orbifold SO(10) GUT model. 
Introducing a singlet chiral superfield $S$, 
 we consider the superpotential%
\footnote{
The renormalizable term, $S (\bar{\phi} \phi)$, 
 can be forbidden by introducing a discrete symmetry \cite{SmoothHI}, 
 for example, $\phi \to -\phi$ and $\bar{\phi} \to \bar{\phi}$. 
}, 
\bea
 W= \lambda 
 S \left( -\mu^2+\frac{(\bar{\phi} \phi)^2}{M_5^2} \right), 
\label{smooth}  
\eea
where $\lambda$ is a dimensionless coefficient, 
 $\mu$ is a dimensionful parameter, and 
 $M_5$ is the 5D Planck mass. 
SUSY vacuum conditions lead to non-zero VEVs for 
 $\langle \phi \rangle = \langle \bar{\phi} \rangle = \sqrt{\mu M}$, 
 by which the PS symmetry is broken down to the SM one, and thus  
\bea 
  v_{\rm PS} = \sqrt{\mu M}.   
\eea 
It is theoretically natural to identify 
 $M_5$ as the GUT scale, $M_5 \sim M_{\rm GUT}$. 
From the analysis of the gauge coupling unification 
 in the context of the 5D orbifold GUT \cite{F-O1}, 
 we found that $v_{\rm PS}= 1.2 \times 10^{16}$ GeV 
 and $M_{\rm GUT} = 4.6 \times 10^{17}$ GeV. 
Independently of the analysis of the gauge coupling unification, 
 it has shown in \cite{F-O2} that 
 this smooth hybrid inflation model, 
 where the inflation trajectory is approximately 
 parameterized by the scalar component of $S$, 
 can reproduce the WMAP data 
 by $v_{\rm PS}=1.2 \times 10^{16}$ GeV and $M_5$ 
 being the the same order of magnitude as $M_{\rm GUT}$.

Now we discuss the main topic of this paper:  
 the non-thermal leptogenesis. 
The relevant part of the superpotential is  
\bea
 W=  \lambda S \left( 
    -\mu^2+\frac{(\bar{\phi} \phi)^2}{M_5^2} \right)
+ \frac{Y_R^{ii}}{M_5} F_{Ri}^c F_{Ri}^c  \left(\phi \phi \right),   
\label{Inf-nuR}
\eea 
where without loss of generality, we work on 
 the mass diagonal basis of the right-handed neutrinos. 
The inflaton which is the scalar component of $S$ 
 couples with the scalar right-handed neutrinos 
 in the scalar potential,  
\bea 
 V \supset 
 \left| \frac{\partial W}{\partial \phi}\right|^2  
 =  \left| 
  \lambda S \frac{2\bar{\phi}(\bar{\phi} \phi)}{M_5^2} 
+ 2 \frac{Y_R^{ii}}{M_5} \tilde{F}_{Ri}^c \tilde{F}_{Ri}^c  \phi 
\right|^2. 
\eea 
Parameterizing the inflaton field $\sigma = \sqrt{2} \Re[S]$, 
 the inflaton mass is found to be 
\bea 
 m_\sigma = 2 \sqrt{2} \lambda \frac{v_{\rm PS}^3}{M_5^2}, 
\eea 
 and the interaction between the inflaton and 
 the scalar right-handed neutrinos 
\bea 
 {\cal L}_{\rm int}  = - 
 \sqrt{2} \lambda \left( \frac{v_{\rm PS}}{M_5}\right)^2 M_i \; \sigma 
   \left( \tilde{F}_{Ri}^c \tilde{F}_{Ri}^c  + {\rm h.c.}
   \right),   
\eea 
 where $M_i = 2 Y_R^{ii} (v_{\rm PS}^2/M_5)$ is mass 
 of the (scalar) right-handed neutrino of the $i$-th generation, 
 and we set $M_1 \leq M_2 \leq M_3$ 
 without loss of generality. 
The partial decay width of the inflaton into the $i$-th  
 generation scalar right-handed neutrino, 
 if kinematically allowed, given by 
\bea
 \Gamma (\sigma \to \tilde{N}_i \tilde{N}_i) 
 = \lambda^2 \frac{M_i^2}{2 \pi m_\sigma} 
 \left( \frac{v_{\rm PS}}{M_5} \right)^4.  
\label{DecayWidth}
\eea
Here $\tilde{N}_i$ denotes the scalar right-handed neutrino  
 in the $i$-th generation. 
Since the inflaton and the superfields, $\phi$ and $\bar{\phi}$, 
 have the same mass, the inflaton cannot decay into 
 the superfields. 

In non-thermal leptogenesis, 
 the inflaton decays into (scalar) right-handed neutrinos 
 and then, the CP-violating decay of the neutrinos 
 generates lepton asymmetry of the universe, 
 which is finally converted into baryon asymmetry 
 via the sphaleron processes. 
The resultant baryon asymmetry of the universe 
 is evaluated as 
\bea
\left(\frac{n_B}{s} \right)
&=& - \frac{10}{31} \times 
 \sum_i \left(\frac{n_{N_i}}{s} \right) \left(\frac{n_L}{n_{N_i}} \right)
 \nonumber\\
&=& - \frac{10}{31} \times \frac{3}{2} 
 \sum_i {\rm BR}(\sigma \to \tilde{N}_i \tilde{N}_i)
\left(\frac{T_R}{m_\sigma} \right) \epsilon_i \;,
\label{YB}
\eea
where the sum is taken to be scalar right-handed neutrinos 
 kinematically allowed, and the CP-violating parameter 
 is given by \cite{epsilon} 
\bea 
\epsilon_i = - \frac{1}{2 \pi (Y_\nu Y_\nu^\dag)_{ii}}
\sum_{j\neq i} \mbox{Im} \left[(Y_\nu Y_\nu^\dag)_{ij}^2 \right]
 f(M_j^2/M_i^2) 
\label{epsilon}
\eea 
 with the Dirac neutrino Yukawa coupling $Y_\nu$ and 
\bea
f(x) &\equiv& 
 \sqrt{x} \; \mbox{ln}\left(\frac{1+x}{x}\right) 
+ 2 \frac{\sqrt{x}}{x - 1}.   
\eea
Here we have assumed that masses of all scalar 
 right-handed neutrinos are greater than the reheating 
 temperature after inflation. 
This assumption is crucial because if a scalar 
 right-handed neutrino is lighter than the reheating 
 temperature, the scenario becomes thermal leptogenesis
 and  the baryon asymmetry produced is not enough 
 for a low reheating temperature.

\begin{figure}[t]
\begin{center}
\includegraphics[scale=1.2]{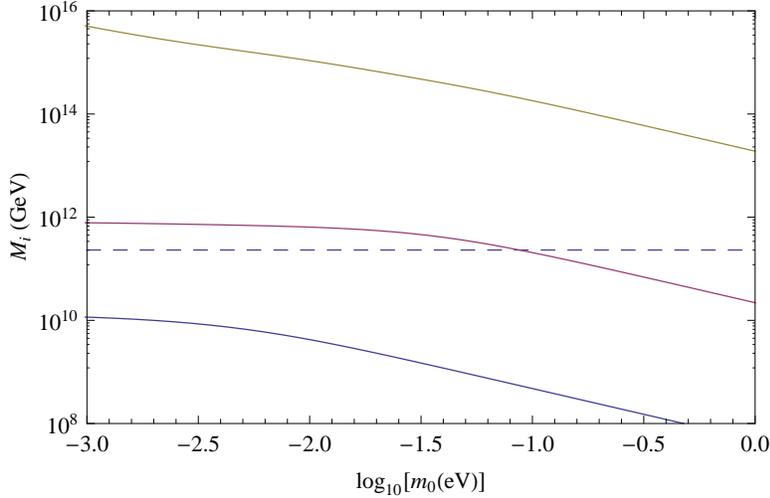}
\caption{ 
The mass spectrum of the scalar right-handed neutrinos 
 as a function of $m_0$ (solid lines) 
 in the normal hierarchical case for the light 
 neutrino mass spectrum. 
The dashed like represents $m_\sigma/2$. 
}
\end{center}
\end{figure}

\begin{figure}[t]
\begin{center}
\includegraphics[scale=1.2]{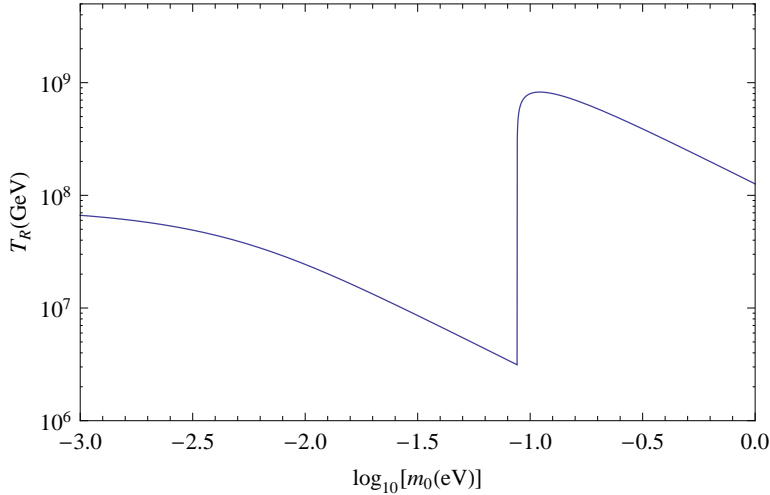}
\caption{
The reheating temperature as a function of $m_0$ 
 in the normal hierarchical case. 
}
\end{center}
\end{figure}

\begin{figure}[t]
\begin{center}
\includegraphics[scale=1.2]{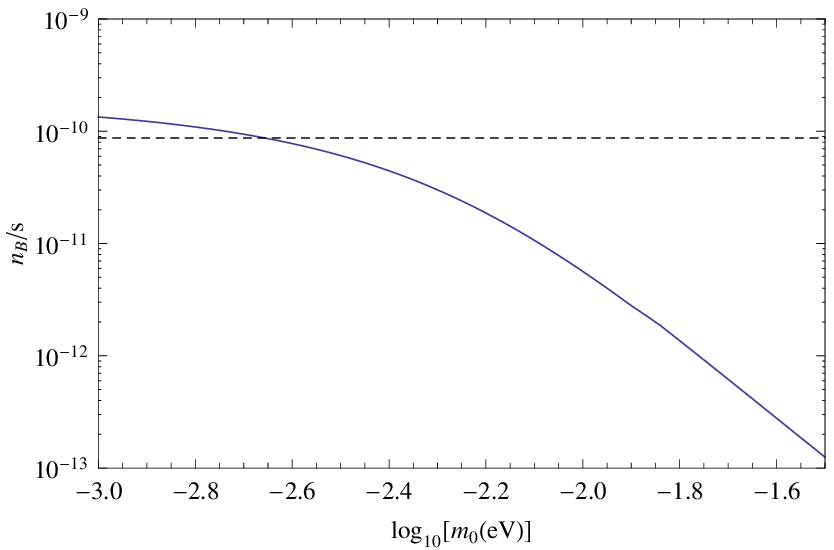}
\caption{ 
The resultant baryon asymmetry produced via 
 non-thermal leptogenesis as a function of $m_0$ 
 (solid line) in the normal hierarchical case. 
The dashed line represents the observed value 
 $Y_B = 0.87 \times 10^{-10}$.  
}
\end{center}
\end{figure}

For the prediction of the resultant baryon asymmetry, 
 we need the information of the Dirac Yukawa coupling, 
 the mass spectra of the scalar right-handed neutrinos 
 and light neutrinos, and the neutrino mixing matrix. 
Through the seesaw mechanism \cite{seesaw}, 
 the light neutrino mass matrix is given by
\bea
 m_\nu=Y_\nu^T M_R^{-1} Y_\nu v_u^2 = U_{MNS} D_\nu U_{MNS}^T 
\eea
in the basis where the mass matrix of charged lepton is diagonal.
Here $v_u$ is the VEV of the up-type Higgs doublet, 
 $M_R$ is the mass matrix of the right-handed neutrinos, 
 and $D_\nu$ is the diagonal mass matrix of light neutrinos. 
In this paper, we consider two typical cases 
 for the light neutrino mass spectrum and 
 describe $D_\nu$ in terms of the lightest mass eigenvalue 
 $m_0$ and the mass squared differences: 
\bea
 D_\nu=\mbox{diag}\left(m_0,~\sqrt{\Delta m_{12}^2 + m_0^2},
 ~\sqrt{\Delta m_{13}^2 + m_0^2}\right) 
\eea 
 for the normal hierarchical case, and
\bea 
 D_\nu=\mbox{diag}\left(\sqrt{\Delta m_{13}^2+m_0^2}, 
 ~\sqrt{\Delta m_{12}^2+\Delta m_{13}^2+m_0^2},~m_0\right)
\eea
 for the inverted hierarchical case. 
Here we adopted the neutrino oscillation data \cite{NuData}: 
\bea 
  \Delta m_{12}^2=7.59\times 10^{-5}~\mbox{eV}^2,
~~\Delta m_{13}^2=2.43\times 10^{-3}~\mbox{eV}^2 
\eea
In addition, we assume the mixing matrix of the so-called 
 tri-bimaximal form \cite{TBM} 
\bea
 U_{MNS} = 
\left(
 \begin{array}{ccc}
 \sqrt{\frac{2}{3}}  & \sqrt{\frac{1}{3}} &  0 \\
-\sqrt{\frac{1}{6}}  & \sqrt{\frac{1}{3}} & \sqrt{\frac{1}{2}} \\ 
-\sqrt{\frac{1}{6}}  & \sqrt{\frac{1}{3}} & -\sqrt{\frac{1}{2}}
      \end{array} \right),
\label{TBM}
\eea
 which is in very good agreement with the current best fit 
 values of the neutrino oscillation data \cite{NuData}. 
As we mentioned above, 
 the data fit for the realistic Dirac mass matrices  
 of the present model is the same as in the minimal SO(10) model, 
 and as an example, we here use the numerical value 
 $Y_\nu$ obtained in \cite{Fuku2} at the GUT scale 
 usual in 4D models $\simeq 10^{16}$ GeV for $\tan \beta=45$:  
\begin{eqnarray}
 Y_\nu = 
\left( 
 \begin{array}{ccc}
-0.000135 - 0.00273 i & 0.00113  + 0.0136 i  & 0.0339   + 0.0580 i  \\ 
 0.00759  + 0.0119 i  & -0.0270   - 0.00419  i  & -0.272    - 0.175   i  \\ 
-0.0280   + 0.00397 i & 0.0635   - 0.0119 i  &  0.491  - 0.526 i 
 \end{array}   \right) .  
\label{Ynu}
\end{eqnarray}    
In this way, we can obtain the (scalar) right-handed neutrino mass 
 matrix as a function of $m_0$, 
\bea 
 M_R = v_u^2 \left( 
  Y_\nu U_{MNS}^*D_\nu^{-1}U_{MNS}^\dagger Y_\nu^T  \right)
\eea 
 with $U_{MNS}$ assumed to be the tri-bimaximal mixing matrix.

\begin{figure}[t]
\begin{center}
\includegraphics[scale=1.2]{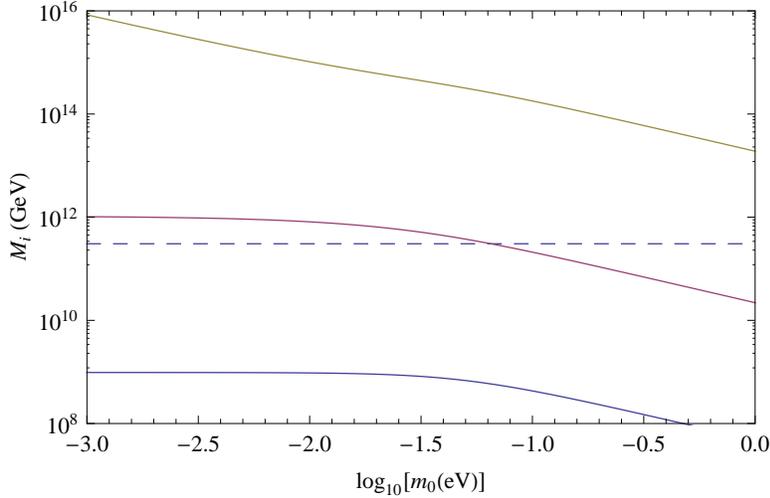}
\caption{ 
The same as Fig.~1 
 but for the inverted hierarchical case. 
}
\end{center}
\end{figure}

\begin{figure}[t]
\begin{center}
\includegraphics[scale=1.2]{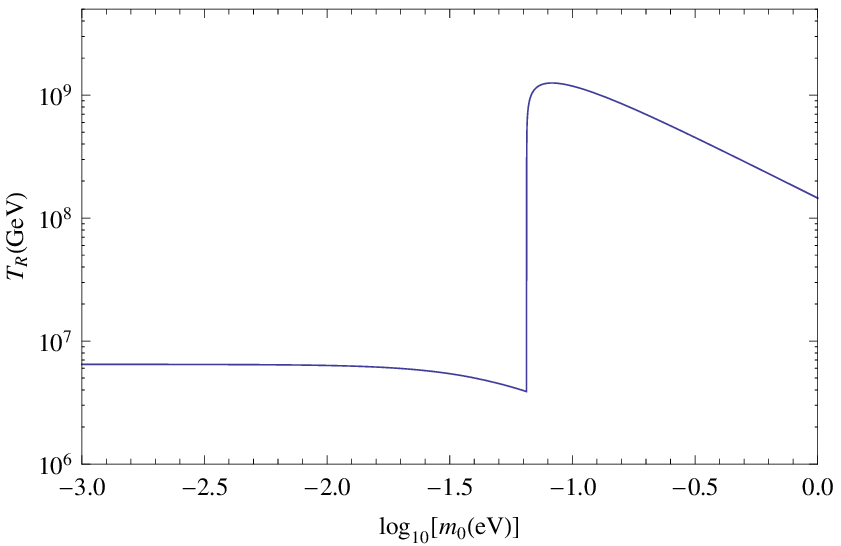}
\caption{
The same as Fig.~2 
 but for the inverted hierarchical case. 
}
\end{center}
\end{figure}

\begin{figure}[t]
\begin{center}
\includegraphics[scale=1.2]{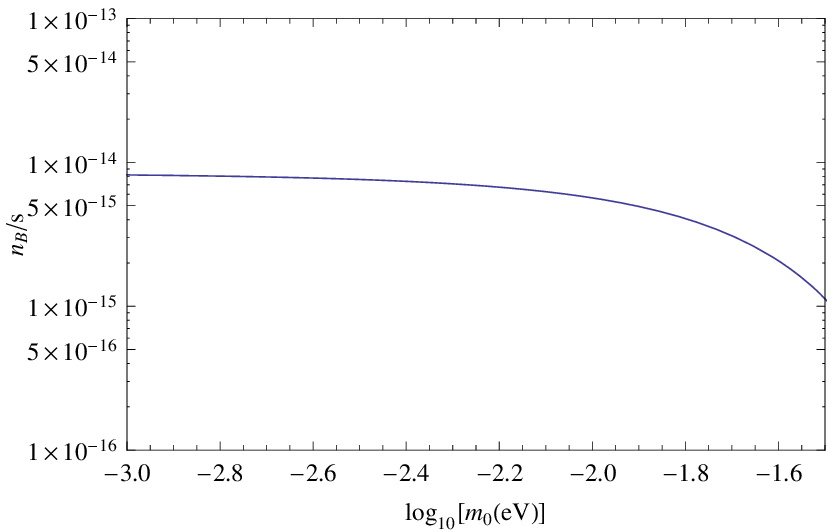}
\caption{ 
The same as Fig.~3 
 but for the inverted hierarchical case. 
}
\end{center}
\end{figure}

Now we are ready to analyze the baryon asymmetry 
 produced in the model. 
Once the parameters in the smooth hybrid inflation, 
 $\lambda$, $v_{\rm PS}$ and $M_5$, are fixed, 
 the baryon asymmetry of the universe of Eq.~(\ref{YB}) 
 is calculated as a function of only $m_0$, 
 providing the reheating temperature evaluated as 
\bea 
 T_R = \left( \frac{45}{4 \pi^3 g_*} \right)^{\frac{1}{4}} 
       \sqrt{\Gamma M_{\rm Pl}},  
\label{TRH}
\eea 
 which is also a function of $m_0$. 
Here $\Gamma$ is the total decay width of the inflaton, 
 $M_{\rm Pl} = 1.2 \times 10^{19}$ GeV, and $g_* \sim 200$.

In our analysis, we take 
 $v_{\rm PS} = 1.2 \times 10^{16}$ GeV 
 and $M_5 = 4.6 \times 10^{17}$ GeV 
 according to the values found in \cite{F-O1} 
 which realize the gauge coupling unification. 
As an example, we fix $\lambda=0.02$. 
For these parameters fixed, Fig.~1 shows 
 the mass spectrum of the scalar right-handed neutrinos 
 as a function of $m_0$, together with 
 $m_\sigma/2$, in the normal hierarchical case.  
For $m_0 \lesssim 0.1$ eV, the inflaton decays 
 into only a pairs of the scalar right-handed neutrinos 
 in the first generation.

The reheating temperature as a function of $m_0$ 
 is depicted in Fig.~2. 
The jump around $m_0 \sim 0.1$ eV is because 
 the decay channel of inflaton into the scalar 
 right-handed neutrino in the second generation 
 is opened up there and the reheating temperature 
 becomes higher than $M_1$, $T_R > M_1$, 
 so that we exclude the region $m_0 \gtrsim 0.1$ eV 
 in our analysis.

In Fig.~2, the reheating temperature exceeds 
 its BBN bound $T_R \lesssim 10^6$. 
However, this bound is not applicable if the gravitino 
 is heavy, $m_{3/2} \gtrsim 100$ TeV, in which case 
 gravitino in the early universe decays before BBN takes place. 
As has been investigated in \cite{F-O3}, 
 the gaugino mediated supersymmetry breaking 
 is naturally incorporated in our 5D SO(10) GUT, 
 where SUSY breaking is assumed to occur 
 on the brane at $y=0$ and the SO(10) gaugino residing in the bulk 
 directly couples with the SUSY breaking sector, 
\bea
 {\cal L} =c_g \delta (y) 
 \int d^2 \theta \; 
  \frac{X}{M_5^2} 
 {\rm tr} \left[ {\cal W}^\alpha {\cal W}_\alpha \right], 
\label{gaugino}
\eea 
 where $X$ is a singlet chiral superfield which breaks SUSY 
 by its F-component VEV ($F_X$), and $c_g$ 
 is a dimensionless constant. 
Then, the gaugino obtains the SUSY breaking soft mass, 
\bea
 m_\lambda =c_g \frac{F_X}{M_5^2}M_c 
  \simeq c_g \frac{F_X}{M_P} \left(\frac{M_5}{M_P}\right) 
  \simeq c_g m_{3/2}\left(\frac{M_5}{M_P}\right), 
\eea
 where the compactification scale $M_c$ comes from 
 the wave function normalization of the bulk gaugino, 
 we have used the relation between the 4D and 5D Planck 
 masses, $M_P^2 \simeq  M_5^3/M_c$ with the reduced Planck mass
 $M_P = 2.4 \times 10^{18}$ GeV,  
 and $m_{3/2} \simeq F_X/M_P$ is gravitino mass. 
In this paper we adopt $c_g \lesssim 0.1$, 
 so that $m_{3/2} \gtrsim 100$ TeV for $m_\lambda \sim 100$ GeV.

Finally, in the normal hierarchical case for the light 
 neutrino mass spectrum, 
 we show the resultant baryon asymmetry of the universe 
 generated via the non-thermal leptogenesis as a function of 
 $m_0$ in Fig.~3, together with the currently observed value 
 \cite{WMAP} 
\bea 
 Y_B =\frac{n_B}{s}=0.87 \times 10^{-10}.  
\eea
We find the observed value is reproduced for 
 $m_0 \simeq 1.8 \times 10^{-3}$ eV.

We repeat the same analysis for the inverted hierarchical case. 
Fig.~4 shows the mass spectrum of the scalar right-handed neutrinos,  
 and the reheating temperature is depicted in Fig.~5. 
The resultant baryon asymmetry is shown in Fig.~6 
 as a function of $m_0$. 
We find that in the inverted hierarchical case 
 the baryon asymmetry produced in non-thermal leptogenesis 
 is too small to be consistent with the observation.

In summary, we have studied the non-thermal leptogenesis 
 in the scheme of 5D orbifold SO(10) GUT 
 with the smooth hybrid inflation. 
With an unambiguously determined Dirac Yukawa couplings 
 and an assumption for the neutrino mixing matrix 
 of the tri-bimaximal from, we have analyzed 
 the baryon asymmetry of the universe via non-thermal leptogenesis 
 in two typical cases for the light neutrino mass spectrum, 
 the normal and inverted hierarchical cases. 
The resultant baryon asymmetry is given as a function 
 of the lightest mass eigenvalue of the light neutrinos $m_0$. 
In the normal hierarchical case, 
 for $m_0 \simeq 1.8 \times 10^{-3}$ eV, 
 the model predicts a suitable amount of the baryon asymmetry 
 through non-thermal leptogenesis, 
 while in the inverted hierarchical case, 
 the predicted asymmetry is too small to be consistent  
 with the observations. 
As can be seen from Eqs.~(\ref{DecayWidth}) and (\ref{TRH}), 
 a mildly small $\lambda$ guarantees $M_1 \gg T_R$ and 
 this is crucial for the realization of non-thermal leptogenesis 
 where we can neglect wash-out processes.

Our 5D orbifold SO(10) GUT was originally constructed 
 in order to remedy problems of the minimal SO(10) GUT 
 in particle physics. 
It is very interesting that the parameters determined 
 from particle physics give the consistent observational 
 values of WMAP coming from quite different origins of cosmology.
Leptogenesis may be placed just in the midst of 
 particle physics and cosmology among others and 
 is very sensitive to the parameter of particle physics. 
Our theory is consistent with it, giving additional 
 constraints on the lightest neutrino mass.

\section*{Acknowledgments}
We are grateful to K.Kohri for useful argument 
 on the gravitino problem. 
The works of T.F. are supported in part by 
 the Grant-in-Aid for Scientific Research from the Ministry 
 of Education, Science and Culture of Japan 
 (\#20540282). 


\end{document}